# Benchmarking Python Tools for Automatic Differentiation

Andrei Turkin, Aung Thu

*Abstract*—In this paper we compare several Python tools for automatic differentiation. In order to assess the difference in performance and precision, the problem of finding the optimal geometrical structure of the cluster with identical atoms is used as follows. First, we compare performance of calculating gradients for the objective function. We showed that the PyADOL-C and PyCppAD tools have much better performance for big clusters than the other ones. Second, we assess precision of these two tools by calculating the difference between the obtained at the optimal configuration gradient norms. We conclude that PyCppAD has the best performance among others, while having almost the same precision as the second-best performing tool – PyADOL-C.

## I. Introduction

In many different applications one who solves numerical computing problems has to deal with the exact derivative calculation task. This includes Jacobians and Hessians calculation which are used for solving ordinary and partial differential equations as well as for finding solutions to different optimization problems. Nowadays one of the possibilities to solve the task is to apply the algorithmic or automatic differentiation technique (see, for instance, [1,2,3]). It should be noted that automatic differentiation is neither numerical nor symbolic differentiation, though the main principle behind the procedure of computing derivatives is partly symbolic and partly numerical [4].

There are two ways to implement automatic differentiation for computer programs: *operator overloading* and *source transformation*. Operator overloading is a technique that implies redefinition of such elementary operation as summation, multiplication, division to update the associated gradient object by means of differentiation rules. Source transformation is another way to implement differentiation which implies rewriting the code so that it contains the implementation of a gradient for the piece of code. While the implementation of a source transformation technique is much more complex than the operator overloaded one, it usually leads to faster run-time speeds [5].

For further speculations on some advantages and disadvantages of these approaches we refer to [6].

Nowadays many researcher use Python as a scientific environment, while applying many third-party open source libraries to computation tasks. Some of them can be used to implement automatic differentiation for Python code. The simplest way to apply automatic differentiation to Python programs is to use one of the following tools: PyADOL-C [7], PyCppAD [8], CasADi[9], Computation Graph Toolkit (CGT) [10], Theano [11,12], or AD [13]. All of these tools can be used for Jacobian evaluation by applying the operator overloading technique to implement automatic differentiation except CasADi, which is based on source code transformation (see issue 884 on the GitHub page for CasADi project for further details). In the next section we briefly describe these tools and provide information about their features while solving a cluster of identical atoms optimization problem which we describe in Section 3.

The purpose of this benchmarking is to call attention to automatic differentiation in Python, to provide information on main features of programming tools, and to highlight the advantages of using each of them. The next section is a brief description of several tools for automatic differentiation that can be used in Python. In section 3 we provide information on the problem we solve with the tools, while speculating on their features. We conclude with the experimental results for the problem.

## II. Automatic Differentiation in Python

In this section we provide information on the automatic differentiation tools in Python and point out some difficulties in using it. First, it is important for any tool to be easy-to-install and to-use i.e. one should use it with no or little modifications of the code to implement automatic differentiation. Second, it should be mentioned, that the main challenge of using automatic differentiation tools in python is the difference in syntax and data initialization for the procedure. In this section we provide information about the tools listed before as well as point out some aspects of their usage. Before discussing some details, it should be mentioned that not all of the tools can be easily installed with the *pip* environment. PyCppAd, CasADi as well as CGT must be installed manually by using the information from the corresponding web-sites while PyADOL-C, Theano and AD may be by using special environment. For instance, Theano and AD can be installed by means of *pip*, while the PyADOL-C tool by using Homebrew in MacOS. For further detailes we refer to the web-sites of these packages.

A. Turkin is with the National Research University of Electronic Technology; Dorodnicyn Computing Center, Federal Research Center "Computer Science and Control" of Russian Academy of Sciences, Russia (e-mail: andrei_turkin@hotmail.com).
Aung Thu is with the National Research University of Electronic Technology (e-mail: aungaungthu61050@gmail.com)

## A. PyADOL-C

ADOL-C [14] is a well-known C++ tool for automatic differentiation which implements it by operator overloading technique. A Python wrapper for it is PyADOL-C that uses the same convenient driver to include automatic differentiation into a Python program by means of the following functions. The *trace_on* and *trace_off* functions mark some code section that is going to be differentiated; the *adouble* function declares active variables i.e. ones to be used for differentiation; the *independent* and *dependent* functions is used to mark independent variables. For more details, we refer to [7], because the functionality of the C++ package and the one in Python almost the same except usage of *adouble* variables.

## B. PyCppAD

The PyCppAD [8] tool is another wrapper for a C++ package which is called CppAD [15]. In order to implement its functionality in Python, it uses the same Boost.Python [16] interface to its C++ implementation as PyADOL-C does. PyCppAD uses *independent* function to mark an independent variable and *adfun* one to mark a section of code which one would like to differentiate. It can be done by using *jacobian* method of the object that *adfun* returns.

## C. CasADi

CasADi is a framework for automatic differentiation and numeric optimization [9,17]. This is a tool with broad functionality which focus is on optimal control. While the other packages use operator overloading technique to implement automatic differentiation, CasADi, in its current form, uses source transformation. As the tools above, CasADi exploits the same approach to provide its functionality as well as efficiency of C++ implementation to Python, but uses SWIG [18] instead of Boost.Python [16] that we mentioned before. CasADi uses special syntax for marking active variables and creating function objects. The later ones can be used for the purpose of automatic differentiation.

## D. Theano

Theano is a large Python library that has tight integration with Numpy as well as possibilities to use GPU for data-intensive calculations. Even though Theano is used mainly in the field of deep learning, it has the differentiation capabilities that seems useful to speculate on in this article. Theano uses special macros from the theano.tensor module to create variables and to calculate derivatives. In order to mark active variables Theano provides a list of data types. For instance, *dvector* returns, as it calls, a symbolic variable for a 1-dimensional *ndarray* with float64 precision. To differentiate some expression Theano uses the macro *grad* from the *theano.tensor* module.

## E. Computation Graph Toolkit

This package replicates Theano functionality for automatic differentiation when increasing computational efficiency. Despite the fact that CGT [10] is underdevelopment we would like to provide some information on it. Computation Graph Toolkin (CGT) uses several functions to define active variables: *scalar*, *vector*, *matrix* as well as the *shared*, which can be used for the same purposes as the ones in Theano.

## F. AD

The tool which is called AD [13] is a Python library that uses *adnumber* function to declare an active variable, which is an object with several methods. These methods are used to perform automatic differentiation. An important feature of this tool is a possibility to calculate gradients and hessians from a Python function by means of the following one: *gh*. Two functions that are returned ones can be used in optimization module of the SciPy package.

It should be noted that all the aforementioned tools may be used only after the code modifications by means of the discussed functions. The only exception is the AD tool that can be applied directly to a Python function that one is going to differentiate by means of *gf*.

## III. EXPERIMENTS AND RESULTS

All the Python tools we mention in section 2 can be used to calculate derivatives of Python functions, however, they are not equally fast. Since in many applications the computation of derivatives is the main task, it is important to use an efficient tool, which has maximal run-time speed. Moreover, it should have such precision of gradient calculation that is almost equal to machine one. Therefore, we would like to assess performance each of the tools as well as their precision. For the purpose of this assessment, we chose a cluster optimization problem [19,20], which can be described as follows. We are looking for a geometrical structure of the cluster with identical atoms, the interaction between which is described by pair potentials. Thus, we have to minimize the following energy function [20]:

$$E(\mathbf{x}) = \sum_{i<j} v\big(\rho(x_i - x_j)\big) \quad (1)$$

For the purpose of exemplification, we are using Lennard-Jones potential as it is in [19] and [20]:

$$v(\rho) = \rho^{-12} - 2\rho^{-6} \quad (2)$$

The results of this section can be replicated by using the code one may found on GitHub[1].

In Figures 1(a) and 1(b) we present the results of our benchmark. All the calculations have been performed on a laptop with a 2.2 GHz Intel Core i7 processor and 16 GB of RAM, running MacOS X (10.10.5). We use 1610 clusters with different number of atoms to get information on performance and precision each of the tools we mentioned before. First, we find average time of gradient calculation to assess performance of the tools. Second, by using predefined clusters[2], we calculate norm of the gradients for each of these clusters to assess precision of the tools. Note, despite the usage of the dataset with optimal cluster configurations, we find the precise optimum by using *optimize.minimize* function of the SciPy package with the L-BFGS[21] method. It is necessary due to rounding that takes place when the configurations are stored to the dataset we use for these experiments.

---

[1] An implementation of all the experiments presented in this paper was published on https://github.com/andreiturkin/Python_ClusterOpt.git
[2] see http://doye.chem.ox.ac.uk/jon/structures/LJ.html

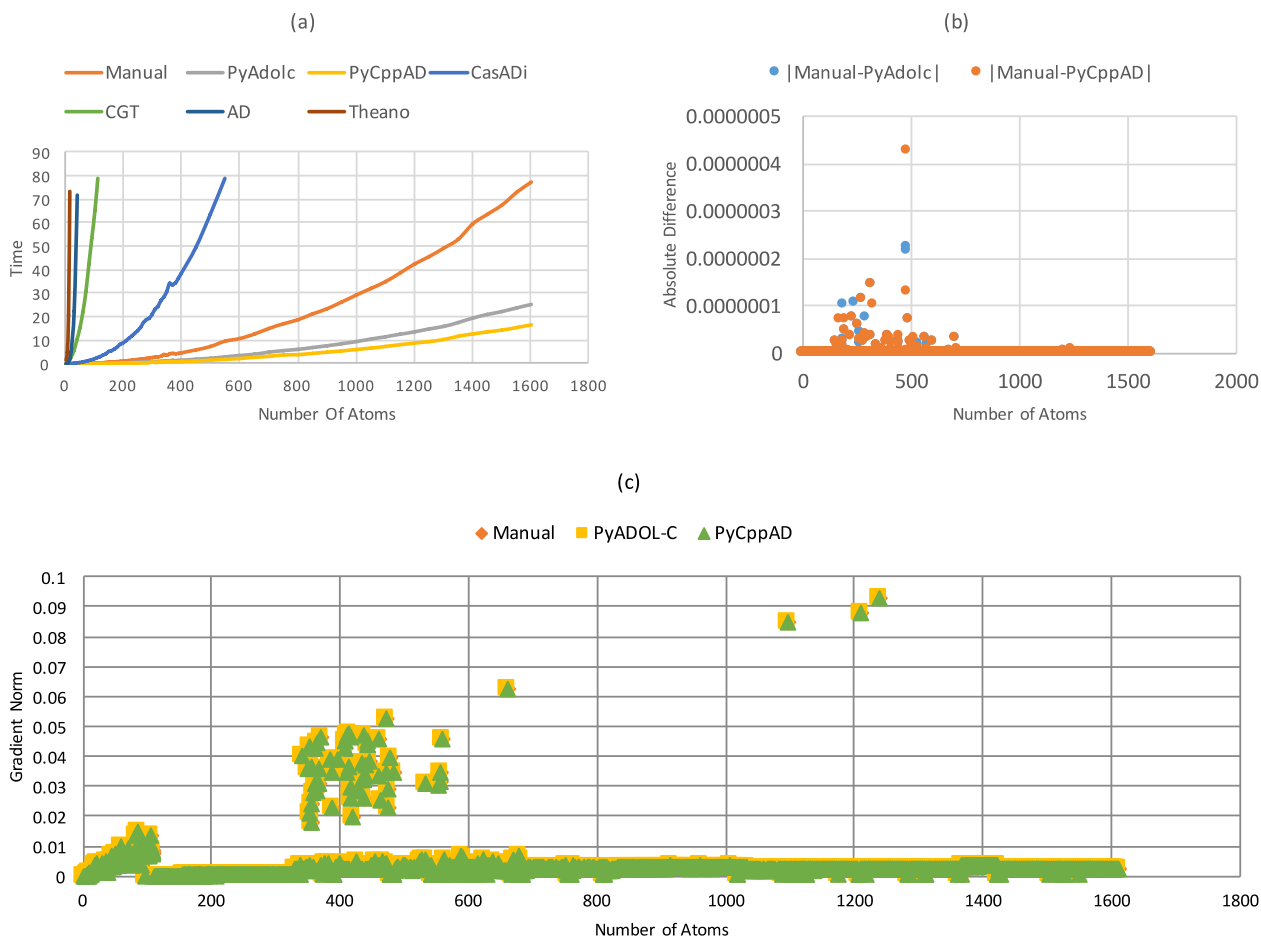

Figure 1. The results of using different tools for automatic differentiation in Python. First, we apply the aforementioned tools to the task of gradient calculation. (a) shows that PyAdolc and PyCppAD have almost the same performance for small clusters, however, for bigger ones the PyCppAD tool is faster. Second, we perform a precision test for the fastest tools from the previous test: PyAdolc and PyCppAD. Since they almost equally close to zero (c) we assess the mean of absolute difference, which is 1.10485E-09 and 1.24189E-09 for PyADOL-C and PyCppAD correspondingly (b); the variances are 1.13743E-16 and 1.77091E-16 for PyADOL-C and PyCppAD.

The optimal solution we use to find out how precise our result is by using norm of the gradient for a cluster.

In Figure 1 we show the average time for gradient calculation obtained by using values from the dataset. [20,22,23,24,25] that consists clusters of up to 1610 atoms. We observe that the PyCppAD-C package shows best results for big clusters, while having almost the same average performance as PyADOL-C. The insufficient performance for clusters with more than 100 atoms as well as the large gap in performance between CGT, Theano, AD, CasADi and the tools we have already mentioned, makes it possible to conclude that it is more convenient to use PyCppAD or PyADOL-C for the purpose of the task.

Note that for all of our tests we use CGT with double precision (precision = string(default=double)) and native backend (backend = option("python","native",default="native")) enabled. We don't use parallel execution graph interpreter, because it makes the results worse.

In Figure 1(b) we provide information on the precision assessment for the following tools, which have the best performance in our previous test: PyCppAD and PyADOLC. Here we exploit the same dataset as in the test before, but use it for the following precision test. First, the initial cluster is obtained from the dataset to calculate gradient norm. Then, we use the L-BFGS algorithm [21] to get the exact location of the minimum by using the following three types of gradients: manually calculated, PyADOL-C and PyCppAD ones. Finally, we calculate the gradient norm at the obtained point by using the same gradients.

We compare the gradient norm values obtained manually with the ones obtained for every cluster by means of the PyADOL-C and PyCppAD tools. They almost equally close to zero (see figure 1(c)); thus it is necessary to assess the absolute difference of two gradient norms: the one that was obtained manually and the one obtained by using some tool. The results show that the automatic differentiation tools give almost the same results: the mean values of the absolute difference is 1.10485E-09 for PyADOL-C and is 1.24189E-09 for PyCppAD; the variances are 1.13743E-16 and 1.77091E-16 for PyADOL-C and PyCppAD correspondingly. Thus, this section can be concluded as follows. PyADOL-C and PyCppAD have almost the same precision of derivative calculation for function 1 with Lennard-Jones potential, however, the PyCppAD tool is distinguishably faster for clusters with more than 600 atoms.

## IV. CONCLUSION

In this paper we have reviewed different Python tools for automatic differentiation and assessed their performance on cluster optimization problem with Lennard-Jones potentials. We showed that PyADOL-C and PyCppAD much faster than the CasADi, CGT, Theano or AD packages. Although the CasADi tool uses source code transformation, the results show that its run-time speed slower than a Python function, which calculates the derivatives manually. Moreover, it performs slower than such tools, which are based on the operator overloading technique, as PyCppAD and PyADOL-C. While both of them have almost the same precision, PyCppAD calculates gradients distinguishably faster for the clusters with more 600 atoms. Therefore, the PyCppAD tool has the best performance for the problem we solve among others, which we use for this benchmarking, while having almost the same precision as PyADOL-C.